\documentclass[nonacm,sigplan,screen,10pt]{acmart}

\renewcommand\footnotetextcopyrightpermission[1]{}
\usepackage{threeparttable}    
\usepackage{multirow}

\usepackage{pifont}

\usepackage{titlesec} \titlespacing*{\section}{0pt}{0.7\baselineskip}
{0.1\baselineskip}
\usepackage{titlesec} \titlespacing*{\subsection}{0pt}{0.5\baselineskip}{0.1\baselineskip}

\titlespacing\section{0pt}{5pt plus 4pt minus 5pt}{1pt plus 2pt minus 1pt}
\titlespacing\subsection{0pt}{4pt plus 4pt minus 3pt}{1pt plus 2pt minus 1pt}
\titlespacing\subsubsection{0pt}{3pt plus 4pt minus 2pt}{1pt plus 2pt minus 1pt}
\usepackage{subfigure}
\usepackage{soul}

\setcopyright{acmlicensed}
\copyrightyear{2018}
\acmYear{2018}
\acmDOI{XXXXXXX.XXXXXXX}
\acmConference[Conference acronym 'XX]{Make sure to enter the correct
  conference title from your rights confirmation emai}{June 03--05,
  2018}{Woodstock, NY}
\acmISBN{978-1-4503-XXXX-X/18/06}
\begin{document}
\title{A Zoned Storage Optimized Flash Cache on~ZNS~SSDs}
\author{Chongzhuo Yang, Chang Guo, Ming Zhao, Zhichao Cao}
\affiliation{%
  \institution{School of Computing and Augmented Intelligence, Arizona State University}
  \city{Tempe}
  \country{USA}
}
\renewcommand{\shortauthors}{Trovato and Tobin, et al.}
\begin{abstract}
Zoned Namespace SSDs (ZNS) are introduced recently to mitigate the block interface penalties of flash-based SSDs. It is a good opportunity for flash cache to address cache throughput and write amplification (WA) issues by fully controlling data allocation and garbage collection via zone-based interfaces. However, there are several critical challenges that need to be addressed including zone-interface compatibility, data management of large zone size, and a better tradeoff between throughput, cache hit ratio, and WA.

In this paper, we present Z-CacheLib, a zoned storage optimized flash cache on ZNS SSDs. In Z-CacheLib, we propose: 1) a new zStorage Engine for ZNS SSDs with low mapping and operational overhead, and 2) a novel zCache Engine with cross-layer optimizations to resolve the throughput regression and WA issues of garbage collection, which consists of delayed data eviction with virtual over-provisioning (vOP), a top-down eviction policy (zLRU) optimized from LRU, and a bottom-up drop mechanism (zDrop) for low WA. Our evaluation shows that Z-CacheLib can achieve up to \textbf{\emph{2X}} throughput, 5\% improvement hit ratio, and \textbf{\emph{almost no WA}} compared to CacheLib with compatible regular SSDs, demonstrating benefits of using ZNS SSDs for cache. Moreover, Z-CacheLib can achieve up to \textbf{\emph{6X}} throughput and \textbf{\emph{92\%}} WA reduction compared with F2FS-based scheme.

\end{abstract}

\settopmatter{printfolios=true}
\maketitle
\pagestyle{plain}

\vspace{-4px}
\section{Introduction}
Flash cache is widely used in today's IT infrastructure to speed up different services such as content delivery network (CDN), storage service, container service, VMs in the cloud, social media applications, and key-value stores \cite{flashcachewiki, li2017pannier, berg2020cachelib,  eisenman2019flashield, mcallister2021kangaroo,tang2015ripq, yang2022cachesack, van2006dynamic, yin2021mapperx, lee2014performance, rocksdb23disaggregating}. Flash cache achieves terabytes-scale of cache capacity with high cost-effectiveness by using flash-based SSDs as the main storage media. Different flash cache systems are introduced to serve different applications and workloads such as dm\_cache at Linux Kernel \cite{van2006dynamic, lee2014performance, yin2021mapperx}, Colossus Flash Cache from Google \cite{holland2013flash, yang2022cachesack}, and CacheLib from Meta \cite{berg2020cachelib, mcallister2021kangaroo}.

Although the flash cache is cheaper and has a much larger capacity than a DRAM-based cache, the lower cache operation throughput and flash endurance issues caused by high write amplification (WA) are the main challenges to be addressed when designing a flash cache \cite{shen2017didacache, berg2020cachelib}.

Since the cache workloads have heavy insertions, updates, and evictions and cause random write of blocks, these non-sequential writes to SSDs cause serious WA due to the device's internal garbage collections (GC) in the Flash Translation Layer (FTL) as well as lower and unstable performance. Regular flash-based SSDs are block devices and cache items need to be packed into a large unit (i.e., 16MiB region) to reduce random writes \cite{berg2020cachelib}. However, the semantic gap between cache management and device management still exists and causes both performance and endurance issues.

Zoned Namespace SSDs (ZNS SSDs) emerged to close the semantic gap between applications and flash devices. We believe the ZNS SSDs will be a better storage device for flash cache than the regular SSDs and open-channel SSDs. ZNS SSDs organize the storage space as large zones (e.g., 96 MiB of Samsung ZNS SSD \cite{ha2023zceph, song2023confzns} and 1077 MiB of Western Digital SSDs \cite{bjorling2021zns}). In each zone, data can only be written sequentially and all the data in a zone will be cleaned together via zone reset. Reads can be performed on these drives just like on regular SSDs, without any limitations. ZNS SSD does not support in-place updates. Therefore, migrating valid data before cleaning the zone (i.e., the GC process) is the responsibility of the upper-layer applications. 

ZNS SSDs represent a new category of flash storage devices, offering management capabilities between traditional SSDs and open-channel SSDs. On one hand, compared with regular SSDs, ZNS SSDs can achieve a higher write performance, lower per GiB cost (less over-provisioning for internal FTL to apply GC), and increased flexibility for applications to design efficient data I/Os and GC mechanisms \cite{bjorling2020zone, bjorling2021zns}. On the other hand, ZNS SSDs are easier and simpler to use and manage than open-channel SSDs since wear-leveling and other management are still maintained by the device \cite{han2021zns+, purandare2022append}. In general, with simpler interfaces, better control of GC, and higher write performance, ZNS SSDs can be better devices for flash cache, effectively addressing performance and flash endurance concerns.

\vspace{4px}
\noindent\textit{\textbf{Challenges.}} However, there are two main challenges that need to be addressed when using ZNS SSDs for flash cache.

\vspace{2px}
1) \textit{\textbf{Efficient Data Management and GC.}} It's challenging to design efficient data management for ZNS SSDs to support flash cache. Existing flash cache systems cannot use ZNS SSDs directly. They do not have zone-compatible interfaces and supports. We need to address zone management and I/O issues including over-provisioning (OP), parallel zone operations, and data mappings. Moreover, the flash cache directly takes the responsibility of GC. Different from the flash cache designs for open-channel SSDs which manage the small blocks (e.g., 8MiB) directly, one ZNS zone can be hundreds of MiB or even GiB in size. Designing a highly efficient GC mechanism for large zones to achieve high cache throughput and low WA is challenging.

2) \textit{\textbf{Tradeoffs in Flash Cache.}} It's challenging to make better tradeoffs between six main criteria of flash cache: cache size, OP ratio, throughput, hit ratio, WA factor, and GC behavior. ZNS SSDs can offer advantages in larger cache size and lower OP ratio. But it's challenging to maintain high throughput and a high hit ratio simultaneously due to its WA and GC issues (we will discuss it in Section \ref{sec:exsiting}).

\vspace{4px}
\noindent\textit{\textbf{Our design.}} To address the aforementioned unresolved issues and challenges, we proposed and developed \textbf{Z-CacheLib}, \textit{a high-performance and flash-friendly persistent cache using ZNS SSDs}. Z-CacheLib has two important modules: 1) a storage engine for ZNS SSDs (called \textit{zStorage Engine}) with low mapping and operational overhead, and 2) a novel \textit{zCache Engine} with cross-layer optimization framework. Storage Engine directly manages the CacheLib I/O unit \textit{region} on ZNS SSDs with parallel writes and low overhead mappings, which resolves the issues of block I/O translation and complex mapping overhead observed in the file system. We further introduce a new cache engine with a cross-layer optimization framework to close the gap between cache and flash devices in CacheLib. In zCache Engine, we propose the novel virtual over-provisioning space (\textbf{vOP}) to motivate and design top-down zone-aware eviction policy (\textbf{zLRU}) and bottom-up GC-aware eviction policy (\textbf{zDrop}). On one hand, Z-CacheLib relies on zLRU to aggressively evict regions based on zone information and reduce GC overhead. On the other hand, with zDrop, Z-CacheLib selectively evicts some of the valid regions during GC to effectively optimize the WA and throughput regression with very little cache hit ratio reduction. 

\vspace{2px}
\noindent\textit{\textbf{Overall results.}}
Z-CacheLib is implemented based on CacheLib and it is open-sourced at Github. We evaluated the throughput, cache hit ratio, and WA of Z-CacheLib with different workloads from CacheBench \cite{cachebench}, and compared it with the other four implementations: CacheLib on compatible regular SSD which has the same hardware as the ZNS SSD (Reg-SSD), CacheLib on ZNS SSDs with a compatible filesystem (ZNS-F2FS), CacheLib on ZNS SSDs with a middle layer (ZNS-Middle), and CacheLib on ZNS SSDs with a simple and direct support (ZNS-Direct). The experiment results show that Z-CacheLib can achieve about 142\% to 519\% cache throughput improvement compared with ZNS-F2FS and up to 108\% higher throughput than regular SSDs with a similar cache hit ratio. The WA factor is nearly 1, significantly lower compared to ZNS-F2FS and ZNS-Middle, which are approximately 2 when writing the same amount of cache data. This difference represents a reduction of up to 50\% in WA factor. Importantly, the cache hit ratio of Z-CacheLib is only about 0.1\% to 0.3\% lower than that of regular SSD and ZNS-F2FS schemes. We also comprehensively evaluate throughput, cache hit ratio, and WA of schemes with different vOP space, OP ratios, and cache sizes. 

\vspace{4px}
\noindent\textit{\textbf{Contributions.}} This paper contributes the following:
\begin{itemize}
  \item We conducted a comprehensive analysis of the three possible designs to adapt CacheLib on ZNS SSDs: F2FS, direct support, and middle-layer schemes, and compared them with CacheLib on compatible regular SSDs. It demonstrates the limitations of using regular SSDs and the zone management issues of using ZNS SSDs in flash cache.
  \item We proposed and designed novel user-space management and GC mechanisms to achieve fast I/O at ZNS SSDs. To achieve high performance and low write amplification, we proposed zLRU and zDrop eviction policies. Further, vOP is introduced to model and motivate cache design using ZNS SSDs. 
  \item Compared to regular SSDs, Z-CacheLib can achieve FIFO-level performance while maintaining an LRU-level hit ratio and almost no WA, showing promising benefits of using ZNS SSDs in persistent flash cache.
  \item To our best knowledge, Z-CacheLib is the first reusable and pluggable flash cache engine designed and optimized for ZNS SSDs. Our project and new eviction policies are designed, implemented, and open-sourced.
\end{itemize}

\section{Background and Related Work}

\subsection{Flash Cache and CacheLib}

In today's IT infrastructure, flash-memory-based cache (flash cache) plays an important role in speeding up the storage systems, applications, VMs, and different data-intensive services \cite{arteaga2016cloudcache, koller2015centaur, li2014nitro, luo2013s, meng2014vcacheshare, li2016cachededup, mcallister2021kangaroo, yang2022cachesack, kgil2006flashcache}. Due to the performance and cost advantages of flash-based SSDs, flash cache is widely used in datacenters to reduce latency and improve throughput \cite{holland2013flash, byan2012mercury, yang2022cachesack, li2014nitro, mcallister2021kangaroo, shen2017didacache} and to speed up the Virtual Machines in cloud \cite{koller2015centaur, arteaga2016cloudcache} and content delivery network (CDN) services for a fast data delivery \cite{zhang2016realizing, kim2015hybrid}. 

However, the cache workload is a typical scenario of intensive non-sequential writes, leading to a fragmented space and high device WA due to the GC applied by FTL \cite{mcallister2021kangaroo}. Making better tradeoffs between flash cache performance, flash endurance (WA), and the cache hit ratio is the main research issue addressed by the previous studies. 1) \textit{Admission policy in flash cache.} Lazy Adaptive Replacement Cache \cite{huang2016improving}, uCache \cite{jiang2013ucache}, Flashield \cite{eisenman2019flashield}, CacheLib \cite{berg2020cachelib}, Kangaroo \cite{mcallister2021kangaroo}, and CacheSack\cite{yang2022cachesack} proposed to utilize DRAM to selectively filter out the seldom accessed blocks for a lower flash writes, which improves both throughput and SSD endurance. 
2) \textit{Workload-based flash cache.}
The photo flash cache at Facebook RIPQ \cite{tang2015ripq}
packs small random writes, co-locates similarly prioritized, and lazily moves updated content to reduce the SSD writes, such that a high cache hit ratio can be achieved with a smaller write amplification. 
CacheDedup \cite{li2016cachededup} introduced deduplication to flash cache to improve both performance and flash endurance by reducing the data writes via deduplication and applying duplication-aware cache replacement algorithms. 3) \textit{Device-based flash cache.} DIDACache \cite{shen2017didacache} uses cache eviction as GC to clean the data in the GC blocks to bridge the gap between key-value cache manager and the underlying flash devices on the open-channel SSD platform. CacheLib recently supported Flexible Data Placement(FDP) over NVMe \cite{fdp} to reduce the device WA significantly even in high SSD utilization.

CacheLib is a general-purpose hybrid caching engine combining DRAM cache and flash cache for applications in different working sets \cite{berg2020cachelib,mcallister2021kangaroo}. It is widely used by IT companies such as Meta, Intel, Databricks, and Nebula \cite{cachemeetup, mcallister2021kangaroo}. All the cache items are inserted into DRAM cache first. When an item is evicted from the DRAM cache, it can be selectively written to the flash cache. To improve performance and reduce WA, CacheLib uses log-structured cache to manage large large objects (larger than 2KB size) on flash. The cache objects are packed into \textit{regions} as the larger basic I/O unit (about 16 MiB). 
CacheLib can handle workloads with highly variable sizes and it has a variety of caching policies (e.g., LRU, Segmented LRU, FIFO, 2Q, and TTL) in DRAM cache and widely used policies (i.e., LRU and FIFO) in flash cache to support different workloads in the production environment.

\subsection{ZNS SSDs}

Zoned Namespace SSD (ZNS SSD) was introduced to achieve better isolation, avoid the unnecessary internal GC overhead, and reduce the per GiB price with the zone-based APIs \cite{bjorling2020zone, bjorling2021zns, han2021zns+}. ZNS SSD divides the storage space into zones (e.g., 96 MiB or 1077 MiB per zone). Similar to existing zoned block devices such as Shingled Magnetic Recording (SMR) drives \cite{aghayev2015skylight}, data can only be written sequentially based on write pointer and the write pointer can be moved to start of the zone by resetting the whole zone. Therefore, the responsibility GC is shifted to upper-layer applications. There are a number of systems and applications that are compatible with zone-based devices like F2FS \cite{zonef2fs} and RocksDB \cite{bjorling2021zns}.

Compared to regular SSDs, ZNS SSDs have several advantages. First, ZNS SSDs have a lower per GiB price since less OP space is needed for internal GC \cite{stavrinos2021don}. Second, the GC and zone-based management capability are offloaded to the upper-layer applications. The performance variability and potential tail latency issues are mitigated \cite{bjorling2021zns}. Third, for some applications with certain I/O characteristics, including LSM-based key-value stores and deduplication indexing, the overhead of host-side GC can be effectively relieved by combining data deletion with GC appropriately \cite{choi2020new, jung2022lifetime, li2022efficient, lee2022compaction, purandare2022append, oh2021efficient}.

\section{Motivations}
\subsection{Six Things Flash Cache Needs to Care} \label{sec:factors-in-flashcache}

\begin{figure*}[!t]
\centering
\includegraphics[width=.918\linewidth]{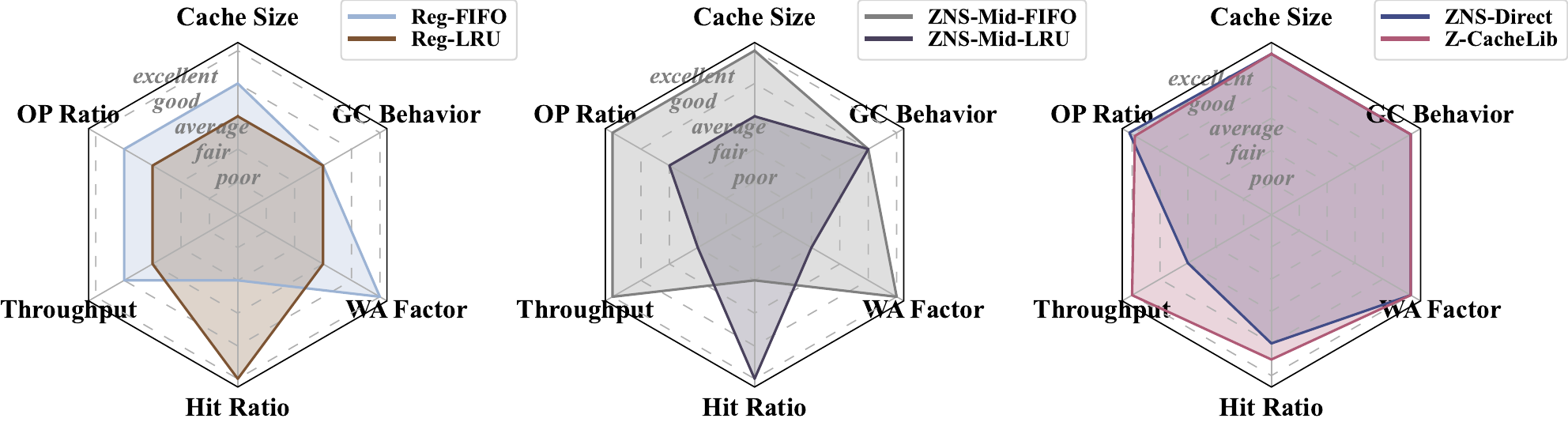}
\caption{The tradeoffs in flash cache under different schemes.}
\label{fig:flash_cache_radar}
\end{figure*}

In this section, we will discuss six main criteria to be considered in flash cache design, including \textit{cache size}, \textit{over-provisioning (OP) ratio}, \textit{throughput}, \textit{hit ratio}, \textit{WA factor}, and \textit{GC behavior}. Figure \ref{fig:flash_cache_radar} shows the tradeoffs under different schemes (discussed in the next sections).

\textbf{Cache size.} The cache size is the amount of space the device provides for the caching system, which is affected by the device space and OP space (cache space = device size - OP space). A larger cache size is desirable for flash cache.

\textbf{OP ratio.} In flash cache, we often allocate a portion of the space as OP to reduce internal GC pressure on the device. Flash-friendly eviction policies require a lower OP ratio, such as the FIFO eviction policy, providing a larger cache size. Therefore, a lower OP ratio is desirable for flash cache.

\textbf{Throughput.} Throughput refers to the number of operations the cache can handle within a specified duration. The higher throughput will increase the effectiveness of the flash cache and optimize the application's overall throughput.

\textbf{Hit ratio.} The hit ratio is the percentage of successful cache lookups. A higher hit ratio can achieve higher overall system performance, especially when there is a large performance gap between the cache and the backend. This factor is often influenced by the cache space and cache policy.

\textbf{WA factor.} Flash requires moving valid data within an erase block to another block during block cleaning to ensure data correctness. The WA factor is the ratio of overall on-device data write (including write caused by GC) and cache engine write. An eviction policy with many random writes may lead to a high WA factor. A large WA factor shortens the device lifespan, so a lower WA factor is preferred.

\textbf{GC behavior.} The GC results in more data being written to the device and occupies the device's bandwidth. In flash cache, we prefer user GC to device GC to avoid unstable latency. The GC-free design is the best scheme.

\vspace{4px}
\noindent\textit{\textbf{It's hard to optimize all factors at the same time!}} Many efforts aim to make tradeoffs among those factors to cater to different workloads \cite{eisenman2019flashield,shen2017didacache,mcallister2021kangaroo}. There are many trade-offs among these factors, and it's very difficult to guarantee that the cache design can perform well in all the factors. For example, if we want to have a larger cache size, the OP ratio should be reduced. However, a lower OP ratio may incur lower throughput or a higher WA factor. Also, if we want to achieve a higher hit ratio, advanced eviction policy often leads to a larger WA factor and lower throughput than the simple FIFO policy \cite{yang2023fifo}.

\subsection{Flash Cache Design Trade-offs on Regular SSDs} 
In the current flash-based caching systems, the regular SSDs are commonly used as the storage backend, which has a block interface consisting of logical blocks (usually 512B or 4KiB in size). In regular SSDs, the internal GC is managed by the device itself, which adds complexity to utilizing cache information to optimize all relevant factors. The GC behavior of Regular SSDs is executed internally and such uncontrollable GC may bring unstable latency \cite{bjorling2021zns}. Further, Regular SSDs will encounter write I/O errors when OP ratio is low due to slow cleaning throughput. In our evaluation of the \textit{reg-wc} workload (see Section \ref{sec:evaluation} for details of workloads), the 960 GiB regular SSD needs at least 73 GiB OP space to avoid I/O errors caused by \texttt{device busy}. The situation can not be improved even under the FIFO policy.

In this section, we will discuss the challenges to balancing these trade-offs by analyzing schemes of regular SSDs with FIFO policy (\textit{Reg-FIFO}) and LRU policy (\textit{Reg-LRU}). The leftmost figure in Figure \ref{fig:flash_cache_radar} shows the overall tradeoffs of Reg-FIFO and Reg-LRU.

\vspace{2px}
\noindent\textit{\textbf{Cache space or OP space.}}
To support the block interface, the internal GC in regular SSDs must relocate logical blocks in the physical victim erase block. These additional writes result in device-level WA, performance reduction, and unstable latency. Besides, a large WA factor will explicitly reduce the lifespan of flash devices. As pointed out in previous work \cite{eisenman2019flashield, mcallister2021kangaroo}, the LRU policy can issue many random writes to flash devices, which further reduces throughput and increases the WA factor. Providing OP space is a commonly used solution to alleviate WA but it will lead to a smaller cache size when the reserved device space is fixed. A reduced OP space increases the overhead of updating data, which can lower the throughput of the cache and overall system. Due to limited device space, it is hard to enlarge both spaces simultaneously \cite{oh2012caching}.

\vspace{2px}
\noindent\textit{\textbf{Flash-friendly or hit ratio.}} FIFO is a flash-friendly caching policy with sequential writing and reclaiming behaviors. The throughput and WA are mainly influenced by how many random writes of eviction policy \cite{yang2023fifo}. As shown in Figure \ref{fig:flash_cache_radar}, Reg-FIFO performs well in both throughput and WA factor. Moreover, FIFO policy needs less OP space than the LRU policy. However, FIFO usually has an explicitly lower hit ratio than LRU in most use cases and is often treated as the baseline for hit ratio analysis. The LRU is the most commonly used caching policy to achieve a higher hit ratio, but it is more complex than FIFO and less friendly for multi-threaded processing \cite{qiu2023frozenhot}. Multiple queues FIFO-based policies (e.g., 3Q-FIFO \cite{yang2023fifo}) may incur additional WA factor due to reinsertion when applying to flash cache even if they only consist of sequential writes. The trade-off exists between optimizing for flash storage and optimizing for hit ratio.
\subsection{It's Time to Abandon the Block Interface!}
Over the last few years, one of the main research focuses on ZNS SSDs is to optimize LSM-based key-value store for ZNS SSDs \cite{bjorling2021zns,choi2020new,jung2022lifetime,lee2022compaction,li2022efficient}. We believe ZNS SSDs are also better storage devices for flash cache than regular SSDs, from performance, cost, and SSD endurance perspectives. First, write throughput is one of the main measurement metrics of flash cache. Compared with regular SSDs, ZNS SSDs can achieve a high and stable write throughput \cite{bjorling2021zns}. Second, without the internal OP reserved for FTL, the per GiB cost of ZNS SSDs can be 10\% to 20\% lower. Large-scale flash cache users like large internet companies (e.g., Meta and Twitter) and cloud services providers (e.g., Amazon, Microsoft, and Google) can achieve better cost-effectiveness. More importantly, using ZNS SSDs in flash cache provides a good opportunity to efficiently address the high GC overhead, which can effectively reduce the WA and mitigate the performance regressions.
In general, ZNS SSDs provide a good opportunity for flash cache to make a better trade-off between cache size, OP ratio, throughput, hit ratio, and WA factor. 
\vspace{2px}
\noindent\textit{\textbf{Three possible solutions on ZNS SSDs.}} Based on the existing studies, there are three possible solutions to support CacheLib on ZNS SSDs. First, we use the ZNS SSD compatible file system F2FS with zone support for CacheLib (called \textit{ZNS-F2FS}). Second, we designed a middle layer between the zone interface and the region interface of CacheLib to achieve the mapping and GC management (called \textit{ZNS-Middle}). Finally, we propose a simple way to use ZNS SSD directly: configuring the region the same as the zone size to directly connect cache management with the zone GC (called \textit{ZNS-Direct}). By default, all schemes will use LRU policy.

\textbf{ZNS-F2FS} One straightforward solution is to use the ZNS-compatible file systems such as F2FS \cite{lee2015f2fs, zonef2fs} to support the I/O from CacheLib under its existing block device abstraction engine. However, F2FS requires a small random writable storage device to store the file system metadata, which limits the use cases and increases the deployment complexity. 

\textbf{ZNS-Middle} Motivated by the middle layer designs in SMR and IMR research \cite{manzanares2016zea, wu2019zonealloy, yao2019geardb, liang2021kvimr, hajkazemi2019track}, we proposed and developed a middle-layer between CacheLib and ZNS SSDs. We maintain a mapping for each region from its logic address in the cache management to its physical address (zone number with offset). When one region in the cache is evicted and overwritten, the old data is marked invalid. In ZNS-Middle, We implemented a background GC that executed the zone cleaning by migrating the valid region to another writable zone. The ZNS-Mid-FIFO and ZNS-Mid-LRU schemes of ZNS-Middle are based on FIFO and LRU policy respectively.

\textbf{ZNS-Direct} DIDACache \cite{shen2017didacache} proposed to integrate cache management and GC on the open channel SSDs platform. It matches the minimum erase block in SSDs as a cache management unit to remove redundant mapping and combine cache eviction and device GC. Motivated by DIDACache, we propose and implement a special CacheLib (called ZNS-Direct) with basic zone management to support ZNS SSDs. Similar to the design in DIDACache, we set region size as the zone size, which can achieve GC-free and avoid extra mapping overhead.

\subsection{Flash Cache Design Trade-offs on ZNS SSDs} \label{sec:exsiting}

In this section, we evaluate and analyze the six factors of the aforementioned three schemes with real ZNS SSDs (WD ZN540 \cite{zn540}) and CacheLib directly on compatible regular SSDs WD SN540 (called Reg-SSD). We found two main challenges of designing persistent cache on ZNS SSDs: 1) the high overhead of large erase units in the transport layer, and 2) the large region of cache management.

\begin{figure}[!t]
  \hspace{-8px}
  \centering 
  \subfigure[\scriptsize{Different schemes.}]{ 
  \label{fig:zns-caching-eval} 
    \includegraphics[width=.52\linewidth]{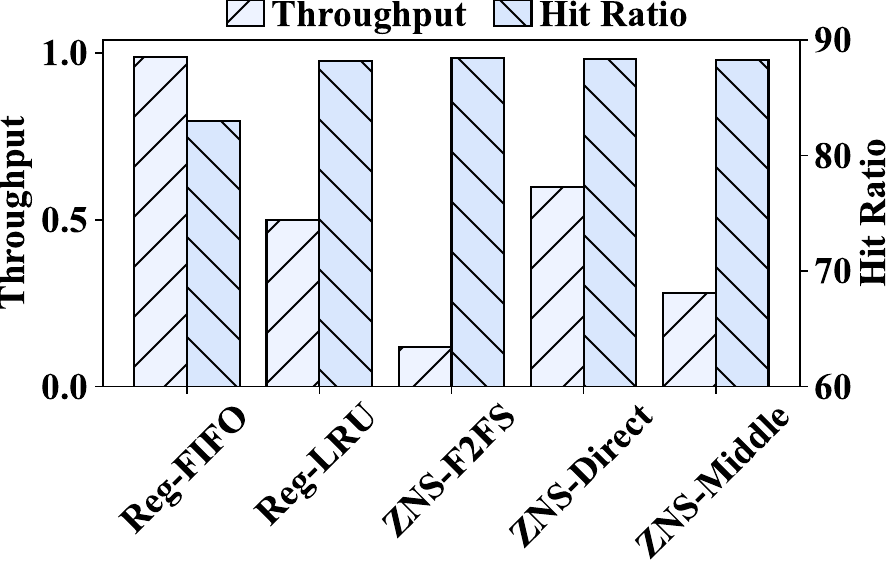}}
  \subfigure[\scriptsize{Different region sizes.}]{ 
  \label{fig:regionsize-ops} 
    \raisebox{4.6mm}{\includegraphics[width=.47\linewidth]{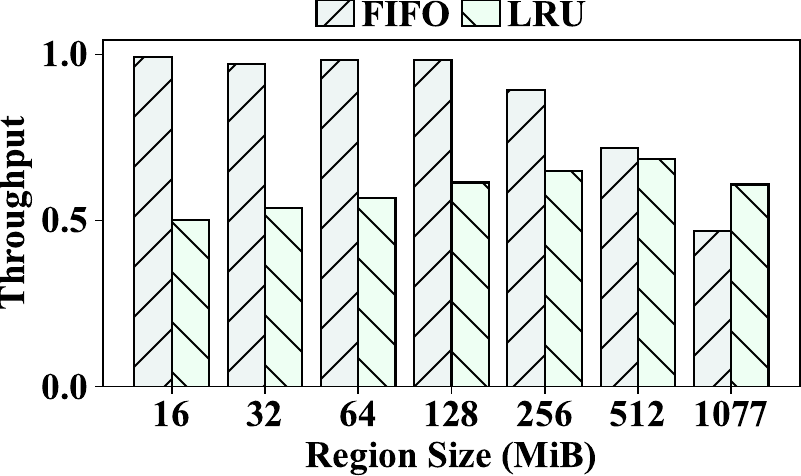}}}
  \caption{Trade-offs on ZNS SSDs.}
  \label{fig:zns caching} 
\hspace{-8px}
\vspace{-4px}
\end{figure}

\vspace{4px}
\noindent\textit{\textbf{Large erase unit in the transport layer.}} We first run Cache-Bench \cite{cachebench} using the \textit{l2-wc} workload and LRU eviction policy (the system setup and workload are presented in Section \ref{sec:evaluation}). Secondly, we conducted evaluations using the FIFO policy on regular SSD (Reg-FIFO) to avoid WA and GC to explore the throughput higher bound of cache schemes. In our evaluation, the cache hit ratios of the four schemes are all at the same level (about 89\%). The throughput variation over time is shown in Figure \ref{fig:zns-caching-eval}. 

The throughput of Reg-LRU during the stable-stage is higher than ZNS-Middle and ZNS-F2FS. The ZNS-Middle can achieve better throughput than ZNS-F2FS but is still lower than Reg-LRU. Compared with ZNS-F2FS, ZNS-Middle is designed specifically for CacheLib instances. It has less indexing and synchronization overhead than F2FS. However, ZNS-Middle still cannot achieve the expected higher throughput than Reg-LRU, and the WA is not reduced (about 2 when CacheLib writes out 1400 GiB data to a ZNS SSD with 800 GiB capacity). The transport layer designs in ZNS-F2FS and ZNS-Middle cause lower performance and fail to reduce WA. Differently, regular SSDs have a smaller erase unit than ZNS SSDs, its internal GC can be more efficient due to less valid data to be migrated, leading to a better throughput. The result also shows that the Reg-FIFO gets pretty good throughput.

\textbf{Observation 1.} \textit{A large erase block size reduces the efficiency of garbage collection when using the middle layer to translate zone interface to block interface}.

\vspace{4px}
\noindent\textit{\textbf{Large region in cache management.}} As shown in Figure \ref{fig:zns-caching-eval}, it is surprising that the GC-free design of ZNS-Direct cannot bring the benefit of higher throughput than Reg-FIFO. To analyze this unexpected behavior of ZNS-Direct, we conducted the experiments using a regular SSD under different region sizes and eviction policies (LRU and FIFO) as shown in Figure \ref{fig:regionsize-ops}.
As shown in Figure \ref{fig:zns-caching-eval} and Figure \ref{fig:regionsize-ops}, ZNS-Direct has similar throughput to using regular SSDs with 1077MiB region and LRU policy. It is reasonable to use regular SSDs to analyze the impact of region sizes. For the LRU eviction policy, the larger region will reduce the random writes, usually leading to higher throughput except for the 1077MiB region size. Note that, the average throughput of FIFO may be lower than LRU due to its unstable and lower hit ratio.

To further investigate the throughput reduction, we conducted evaluations using the FIFO policy to avoid WA. The results show that the throughput will drop as region size increases when the region size is larger than 128 MiB. When the region is large, achieving effective parallel control is challenging. In our evaluation, CacheLib needs 7ms to fill a 16 MiB region in DRAM when the cache is full and region eviction has started. However, CacheLib needs 4s to fill a 1 GiB region, which is pretty slow than 7ms $\times$ 64 = 448ms (i.e., the time overhead is not linearly increasing).

In ZNS-Direct, we set the region size to the zone size which is much larger than the suggested region size (i.e. under 256 MiB in CacheLib). As described in CacheLib, a very large region size can lead to long allocation time and overhead. ZNS-Direct can reduce WA to 1.0 as expected (no GC needed), but the throughput of ZNS-Direct can not outperform the Reg-SSD. Therefore, to achieve a higher performance than Reg-SSD, we need to redesign a zone-compactible engine for CacheLib and achieve fine-grained control during the GC.
\textbf{Observation 2.} \textit{A large cache management unit on flash reduces the efficiency of parallel writing, leading to lower throughput to ZNS-Direct}.

\begin{figure}
\centering
\includegraphics[width=.918\linewidth]{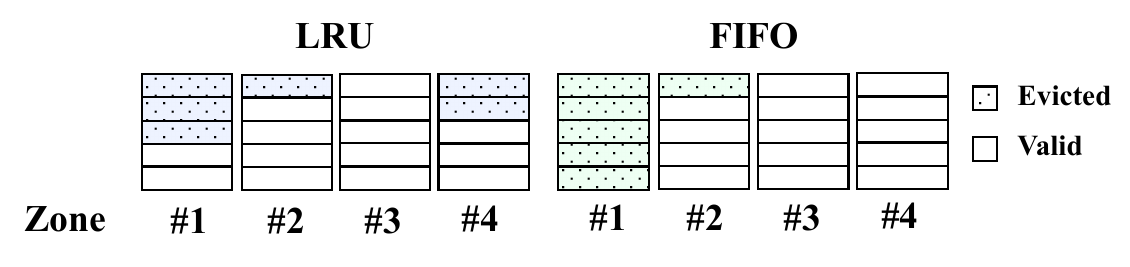}
\caption{Cache space and OP space.}
\label{fig:op-space}
\end{figure}

\vspace{4px}
\noindent\textit{\textbf{Inefficient OP space between cache and device.}} The OP space in the flash cache is utilized to minimize data movements during zone cleaning. The flash cache will write only the cache size amount of valid data into flash, we can consider the cache space as the storage for valid regions. Therefore, the OP space can be viewed as storage for evicted/invalid regions. Therefore, the size of OP space will be equal to the data size of the evicted regions. As shown in Figure \ref{fig:op-space}, in the LRU policy, all evicted regions are distributed across different zones by their recency, collectively building the OP space. In the FIFO policy, evicted regions are sequentially packed into the last few zones.

In ZNS-Middle, when the region is evicted and rewritten, its corresponding data will be marked as invalid for GC to clean. The GC will select one zone with the least valid data as the victim zone to clean. A larger number of valid regions in victim zone will lead to a higher WA factor. For example in Figure \ref{fig:op-space}, FIFO and LRU will both select zone 1 as victim zone. For FIFO, there is very little or even no valid region in the zone 1, leading to lower WA factor. LRU has more valid regions in zone 1 and higher WA factor. This is because the OP space of LRU is not efficiently utilized. For example, the space occupied by evicted regions in zone 2 can not be used until this zone is cleaned. Note that, the evicted region can also contain cache items so it can be accessed by cache to improve space utilization. Therefore, it is beneficial to use these invalid spaces by delaying the real eviction process (i.e., only mark the region as evictable but the region is still valid for cache to be accessed).

\textbf{Observation 3.} \textit{The OP space in the flash cache is not efficiently utilized and the real deletion of the evicted region can be delayed}.

\subsection{It's Time to Redesign an Eviction Policy!}

Different from using ZNS SSDs in the storage systems in which GC must migrate all the valid data to other zones before cleaning the whole zone, evicting any valid data from the cache only lowers the cache hit ratio and does not cause data correctness and corruption problems. Also, compared with the TB-level cache size, evicting some of the cold valid data during GC has very little influence on the cache hit ratio. Therefore, ZNS SSDs provide a good opportunity for flash cache to make a better tradeoff between cache throughput, cache hit ratio, and WA by selectively migrating valid data during GC.

However, there are several challenges that need to be addressed when designing the new engine to achieve both higher throughput and lower WA. \textbf{First}, the block interface in the flash cache and the ZNS-compatible engine should be removed. Instead of maintaining both region and block for mapping and indexing, we can directly use the region as the indexing unit. It can minimize the mapping and operational overhead. \textbf{Second}, a fine-grained mapping and locking mechanism is needed to relieve the performance penalty during metadata operations for better concurrency control. \textbf{Third}, to lower the space cost, a new OP scheme is essential to reserve less space while guaranteeing the overall performance. \textbf{Finally}, a new cache eviction policy should be co-designed with GC to reduce the amount of data being migrated. However, identifying valid data that has little influence on the cache hit ratio but contributes more to the throughput improvement and WA reduction is challenging during GC. As GC is executed by the lower layer storage engine and it has little information about the hotness of cache data.

\section{Z-CacheLib}
To address the aforementioned issues and challenges, we propose a novel zoned storage optimized flash cache on ZNS SSDs, called \textbf{Z-CacheLib}. First, we designed a new storage engine for CacheLib skipping the block interfaces, called \textbf{zStorage Engine} optimized for ZNS SSDs. We further propose the \textbf{zCache Engine} with cross-layer optimizations (presented in Section \ref{sec:zcachelib}) between cache management and storage engine to achieve a better tradeoff between cache performance, WA, and cache hit.
\section{The zCache Engine in Z-CacheLib} \label{sec:zcachelib}

In this section, we introduce the zCache Engine with cross-layer optimizations in Z-CacheLib as shown in Figure \ref{fig:zlru-zdrop}. zCache Engine addresses the gap between cache management and zone management, enabling Z-CacheLib to achieve a better tradeoff between throughput, cache hit ratio, WA.

We first introduce the virtual over-provisioning (\textbf{vOP}) space to provide more OP space. Based on vOP and LRU policy, we propose a novel top-down and zone-aware eviction policy (named \textbf{zLRU}) for Z-CacheLib. The zLRU aggressively evicts the region that only includes little valid data, which significantly improves the throughput and reduces the WA. On the other hand, we also propose a bottom-up eviction policy (named \textbf{zDrop}), which leverages vOP to evict regions during GC. To reduce WA and make vOP like the real OP space, zDrop utilizes the GC information in the victim zone. Because vOP space can be explicitly larger than OP space, the GC overhead can also be significantly reduced. By combining zLRU and zDrop, our new eviction policy for ZNS SSDs can make Z-CacheLib almost GC-free with higher throughput, very low WA, and almost no cache hit ratio regression.

\subsection{Virtual Over-Provisioning Space}

\begin{figure}[t]
\centering
\includegraphics[width=.918\linewidth]{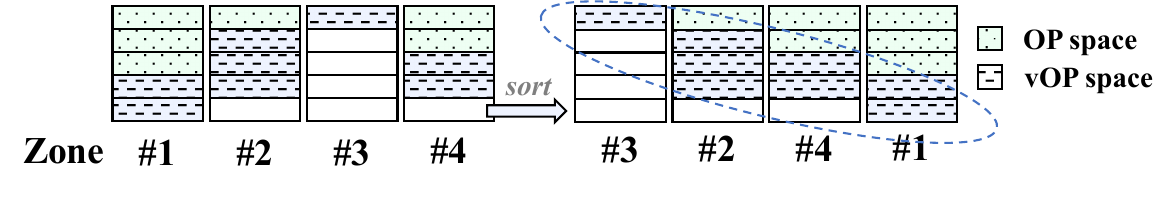}
\caption{The vOP space and OP space.}
\label{fig:vop-space}
\end{figure}

Originally, the zStorage Engine cache engine in CacheLib only had data exchange to read and write the regions. As we found in Section \ref{sec:exsiting}, the OP space in the flash cache is not efficiently utilized and the real deletion of the evicted region can be delayed due to the gap between cache management and device management. To close the gap, Z-CacheLib enables the cache management layer of CacheLib to collect information about physical zones.

In this section, the virtual over-provisioning (\textbf{vOP}) concept is introduced. The vOP space is an additional space provided to the device as shown in Figure \ref{fig:vop-space}. To better understand vOP, we sorted regions by valid data size, and the vOP space refers to the circled area in the figure. The regions in vOP space are evictable but not actually deleted from cache (i.e., the regions are still valid for cache to be accessed). Therefore, the regions in vOP can be accessed and prompted. As discussed in Section \ref{sec:factors-in-flashcache}, there is a tradeoff between OP space and cache size: increasing OP space reduces cache size. However, since the vOP space can be viewed as a part of the cache space, increasing vOP space does not influence the cache size.

\subsection{Top-down eviction }

By leveraging the concept of vOP space, a zone-aware top-down eviction policy (named \textbf{zLRU}) is introduced to achieve early deletion of the regions in the vOP space. The zLRU is motivated by REF \cite{seo2008recently} in buffer management of flash device. The main idea is to reorder the LRU list in vOP space to effectively utilize the OP space (i.e., minimize the amount of valid data in the victim zone as much as possible).

The zLRU is designed based on LRU policy with two new partitions at the LRU list: a regular LRU partition and a vOP partition. In the regular LRU partition, we operate the cache items the same as the original LRU design, and items will be turned into vOP partition when LRU part is full (i.e., LRU partition eviction happens). One item in vOP partition will be promoted to LRU partition when it is accessed. 

\begin{figure}[!t]
\centering
\includegraphics[width=.86\linewidth]{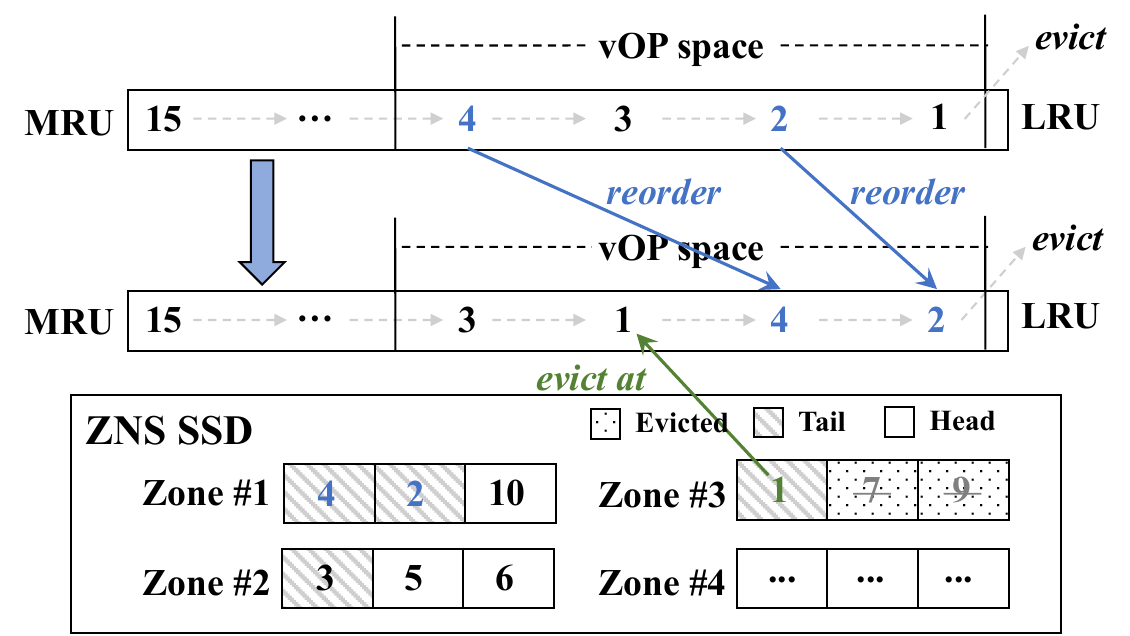}
\caption{The eviction policy of Z-CacheLib.}
\label{fig:zlru-zdrop}
\end{figure}

We collect the number of valid cache regions (excluding regions in vOP and evicted regions) for each zone. When the number of valid cache regions falls below the average number of valid cache regions, we assume that the zone is likely to be evicted soon and mark the zone as a \textit{candidate} for eviction.
To reduce valid data in the victim zone, we proactively move regions whose corresponding zone is an eviction candidate to the end of the vOP partition. For example in Figure \ref{fig:zlru-zdrop}, zone 1 only contains one valid region (region 10) in cache space and it is an eviction candidate, zLRU reorders its regions (4 and 2) to the tail of the zLRU list.

\subsection{Bottom-up eviction} \label{subsec:bottomup}

Top-down eviction cannot guarantee to reclaiming enough free space for incoming new regions. 
Therefore, we introduced the bottom-up GC-aware eviction policy (named \textbf{zDrop}) as a complimentary scheme, which will only be triggered when the free space is lower than the watermark. The main idea is to use region and zone information to delay the real eviction process. Applying bottom-up eviction to regular SSDs is challenging due to the absence of user space FTL and GC mechanisms. This highlights the distinct benefits of ZNS SSDs for caching.

During the GC process, the GC thread identifies the valid regions and propagates the region and zone information up to zCache Engine of CacheLib. The GC process will treat regions in vOP as invalid data. When the victim zone contains regions in vOP, zCache Engine will evict these vOP regions thus their statuses will change from delayed eviction to true eviction. Then, move valid regions in victim zone to another zone and finally reset the victim zone.

One big potential of using ZNS SSDs for persistent cache has not been explored: evicting any valid region will not cause data correctness or corruption issues. Instead of strictly migrating every valid region to a new zone during GC, if the system can identify the valid regions that have a very low influence on the cache hit ratio and drop these regions during GC, the performance influence of GC and WA can be effectively reduced. Therefore, we can set vOP space extremely large for bottom-up eviction, leading to an aggressive zDrop policy (called zDrop-100). In zDrop-100, the victim zone will evict almost all valid regions to ensure almost GC-free. Z-CacheLib will use zDrop-100 policy as the default zDrop policy.

The bottom-up eviction targets at relieving the CacheLib throughput regression during GC and reducing WA. By actively evicting some of the cold valid regions that have little cache hit ratio influence, the amount of data being read and written during GC is reduced significantly.

\subsection{Overview of zStorage Engine}

 zStorage Engine is an I/O engine layer similar to the block device abstraction specifically designed for ZNS SSDs. The original cache management layer of CacheLib can directly use the region as the basic I/Os provided by zStorage Engine. zStorage Engine consists of the data management module and GC module. The data management module maintains the zone status and data mapping and processes the region read and write operations from CacheLib. The GC module selects zones to be cleaned, migrates the valid data, and updates the mapping as background jobs. Further, the concurrency control in zStorage Engine will be discussed in Section \ref{sec:parallel}.
\subsection{Data Management} \label{sec:data-management}

ZNS SSDs only provide zone-based read, write, reset, and reporting APIs, it is essential to maintain the metadata information for each zone during the running time, including zone-metadata to indicate its status, write pointer information, and essential statistics to help the GC.

\textbf{Zone Group.} The zones provisioned for Z-CacheLib are classified into three statuses: \textit{empty zones}, \textit{write zones}, and \textit{read zones}. When the device is clean or when we reset the write pointer, the zones become empty zones. When an empty zone is selected to write the data, it becomes a write zone. We can continue to append data to the write zone until it is full and closed. After a zone becomes full and not writable, it is transferred into a read zone. Only read zones can be cleaned. Users can configure the limit of concurrent write zones, which should be always lower than the open zone limit of the ZNS device. The zone status switching is atomic to ensure correctness and consistency.

\textbf{I/O Units.} To reduce the cost of mapping and metadata operations, instead of splitting the region into blocks (e.g., like the designs in ZNS-F2FS or ZNS-Middle), zStorage Engine uses one mapping entry per region. Region-based data management can also make it easier to achieve a more efficient data allocation and GC with the help of Z-CacheLib cache management layer, which is introduced in Section \ref{sec:zcachelib}.

\textbf{Mapping.}
When we initialize CacheLib, it creates a consecutive virtual space based on the configured cache size and each region has its own virtual address. For example, if we use the FileDevice engine which stores data in a file, the virtual space is mapped to one large file. In zStorage Engine, the data management module translates the virtual address into a zone-based physical address. We maintain a bi-directional mapping between the region virtual address and the ZNS zone-based physical address, called \textbf{zMap}.

The zMap contains two mapping tables, one maps the region virtual address to the ZNS physical address and another one is the reverse mapping (i.e., from ZNS physical address to the region virtual address). Any mapping updates need to update the two mapping tables atomically. For example, when we write the second region (virtual address at 131,072) to the beginning of the second zone (address: 1,048,576). We will update the two mapping data pairs (131,072, 1,048,576) and (1,048,576, 131,072) in zMap atomically.

When writing a new region, Z-CacheLib will find an available region slot in the virtual space and append the data directly based on the virtual address. Then, the data management module selects a write zone with enough free space and appends the region to a certain zone. When the region append is successful, zMap is updated with the new ZNS physical address in the two mapping tables atomically. When reading a region, the virtual space address is searched in the zMap, and the corresponding physical address is returned. 

\subsection{Garbage Collection}

One of the main differences between ZNS SSDs and regular SSDs is the GC process. Regular SSD relies on the FTL to execute the internal GC. Therefore, applications can overwrite the data blocks directly as CacheLib does for regions. In contrast, applications need to implement their own host-based GC when using ZNS SSDs. GC is one of the most important modules of zStorage Engine, which ensures that regions can be written without write stalls after all the zones are fully utilized. Designing a highly efficient GC is challenging since it can influence the foreground primary throughput explicitly as we observed in both ZNS-F2FS and ZNS-Middle. Since GC needs to read and update zMap, the data consistency protection mechanisms will introduce performance overhead. On the other hand, GC occupies the storage bandwidth, which will influence the foreground I/O throughput. 

Since the primary objective of GC is to reclaim the space for upcoming data, selecting a zone with the most invalid data can provide the largest free space in one cleaning cycle, migrate fewer data during GC, and reduce the write amplification (i.e., benefit the SSD lifetime). Therefore, when GC is triggered, it will select one of the read zones with the lowest valid data ratio.

To avoid the slowdown or even write stall before or during GC, GC cannot be triggered after all the free zones are used. Therefore, we use high- and low- watermarks to trigger and stop the GC. The watermark configuration is closely related to the cache size (i.e., the number of reserved zones for Z-CacheLib). For example, the reserved zone number is $Z$, when the empty zone number is fewer than $W_{low}$\% of $Z$, background GC is triggered. When the empty zone number is larger than $W_{high}$\% of $Z$, GC is stopped. In this paper, $W_{low}$ is set to 1 and $W_{high}$ is set to 3. Investigating the way of optimizing the watermarks will be our future work.

\subsection{Over Provisioning}
When reserving the ZNS zones, we need to over-provision several extra zones for GC. We use the following algorithm to estimate the OP ratio lower bound. Suppose the OP ratio is $r_{op}$ (the extra reserved space divided by the virtual space size), the write rate from CacheLib is $T_{cache}$, and the background GC cleaning throughput is $T_{gc}$. The average ratio of invalid data in all zones is $r_{invalid} = r_{op}/ (1 + r_{op})$. The zone to be cleaned has the most invalid data and we assume the ratio of the zone is $r_{clean} = k \cdot r_{invalid}$, while $k \geq 1$. In order to avoid write slowdown or even write stall, the empty space reclaiming rate should be higher than the CacheLib writes rate, $k \cdot r_{invalid} \cdot T_{gc} \ge T_{cache} $. We can get $r_{op} \geq T_{cache} / (k \cdot T_{gc} - T_{cache})$. In our evaluation, $T_{cache} \approx 200$MiB/s, $T_{gc} \approx 600$MiB/s, and $k = 6$. Therefore, the OP ratio is $r_{op} \approx 5.8\%$. Compared with the 10\% to 20\% internal OP ratio of regular SSDs, we can save more space.

\section{Concurrency Control} \label{sec:parallel}
To support high performance under parallel I/Os, we should carefully design the concurrency control on the management of ZNS SSDs.

\textbf{Zone State.} As we mentioned in Section \ref{sec:data-management}, physical zones are divided into three groups: write zone, read zone, and empty zone. In our implementation, we should ensure that the zone status is safe under parallel I/Os. In zStorage Engine, we set the limit of the minimal write zones (soft limitation) to get a better writing performance. We also set a maximum write zone limit (hard limitation) to fit the mandatory max open zones limitation of ZNS SSDs. When the number of write zones reaches the limitation of max open zones, the writes should wait until one open zone is closed and finished (i.e., one write zone is full).

\textbf{I/O Control.} We introduce the parallel writing support to Storage Engine to achieve a better write performance during region write and GC. We maintain multiple write zones and multiple threads can write to different zones concurrently. For each region write, we should select one write zone and allocate one region to it. After one zone is selected for writing, it will not be selected again to ensure only one thread is writing the zone. To ensure data consistency under parallel writing, we use the zone state to avoid the locking mechanism based on each zone. Read operations will only read data on read zones and the written part of write zones, so reads do not need to lock even though the cache is writing the zone. In our implementation, region writes are protected by zone state rather than zone lock, improving the parallel performance. However, a shared read lock is used to avoid data errors when the zone is reset by GC.

\textbf{Data Mapping.}
The data mapping manages the mapping between region id and physical offset. For each mapping access, zMap is protected by the zMap lock. Also, we always follow the data-metadata writing order to ensure consistency. The zMap will be updated once data is successfully written to the device, and it will be searched when reading data from regions. Additionally, it will be changed as data is moved to a new zone during garbage collection.

\textbf{Garbage Collection.} 
To achieve better GC performance, we propose the following background GC process with a fine-grained locking mechanism: 1) The GC module locks the zone state lock and finds a victim zone. 2) Read all valid regions in the victim zone and release the zone state lock. 3) The GC module moves valid regions to a new zone using sequential order in victim zone. 4) If the region is written in the new zone and the region has not been invalided by cache, we update the zMap to ensure data correctness. 5) But the mapping will not be updated when GC module finds the region is evicted while moving data. 6) Repeat steps 3 to 5 until all the regions of this zone are processed. 7) Reset the zone write pointer and update the zone metadata including the status and statistics.

\textbf{Bottom-up Eviction.} Originally, one region could only be evicted by one thread, but the GC module in Z-CacheLib will also evict region using bottom-up eviction policy. So we should pay attention to managing these two different eviction processes. In CacheLib, the region status can be divided into three statuses: evicting, evicted, and flushed. First, if the region is evicting by cache, the bottom-up eviction should wait until the region is fully evicted. To be noted, eviction should read data on ZNS SSDs so the GC should not reset the zone and must wait for the cache eviction. Second, if the region is evicted, the bottom-up eviction can simply skip this region as it is already evicted by cache eviction. Third, if the region is flushed and we find its physical address searched using zMap is not in the current victim zone, it means the region is already flushed again and the current data on victim zone is the old data. Last is the normal situation that the region is flushed and the current data on victim is correct, then we evict the region.

\section{Evaluations} \label{sec:evaluation}
We first present the prototype implementation details of the zStorage Engine and zCache Engine with cross-layer optimizations in Z-CacheLib. Then, we conduct comprehensive evaluations of Z-CacheLib, CacheLib on compatible regular SSD, and CacheLib on F2FS, ZNS-Middle, and ZNS-Direct.

\begin{figure*}[!t]
  \centering 
  \subfigure[\scriptsize{Throughput of \emph{\textbf{l2\_wc}}.}]{ 
  \label{fig:l2-wc} 
    \includegraphics[width=.32\linewidth]{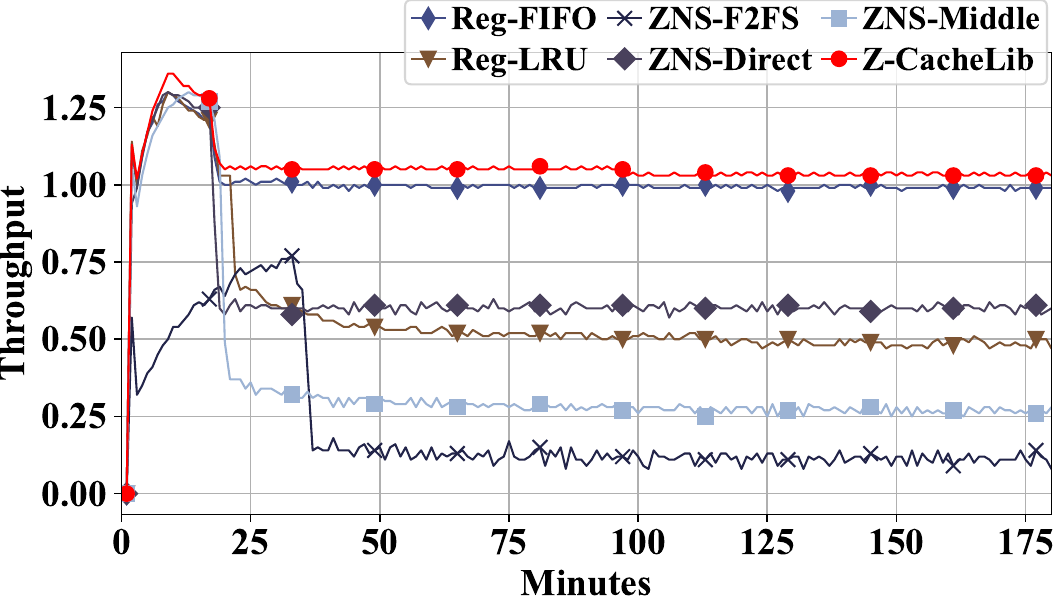}}
  \subfigure[\scriptsize{Throughput of \emph{\textbf{l2\_reg}}.}]{ 
  \label{fig:l2-reg} 
    \includegraphics[width=.32\linewidth]{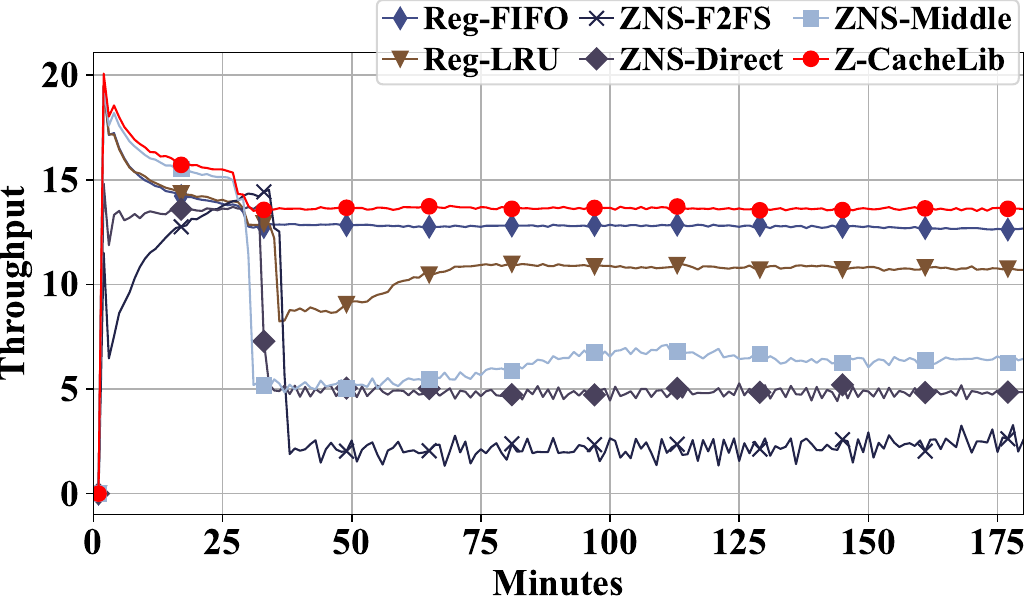}}
  \subfigure[\scriptsize{Throughput of \emph{\textbf{flat}}.}]{ 
  \label{fig:kvcache-reg} 
    \includegraphics[width=.32\linewidth]{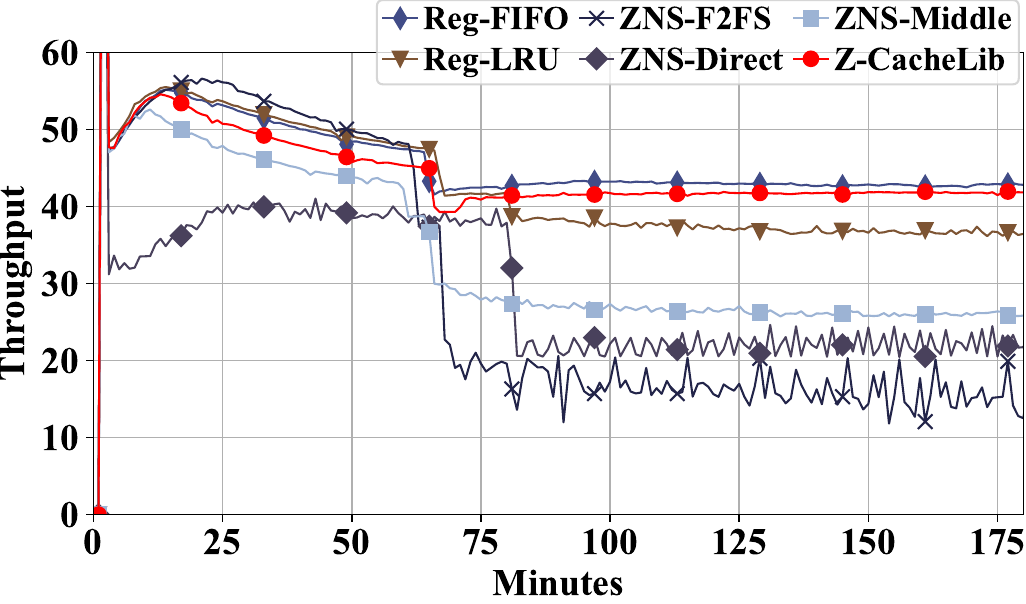}}

  \subfigure[\scriptsize{Cache hit ratio of \emph{\textbf{l2\_wc}}.}]{ 
  \label{fig:l2-wc-ratio} 
    \includegraphics[width=.32\linewidth]{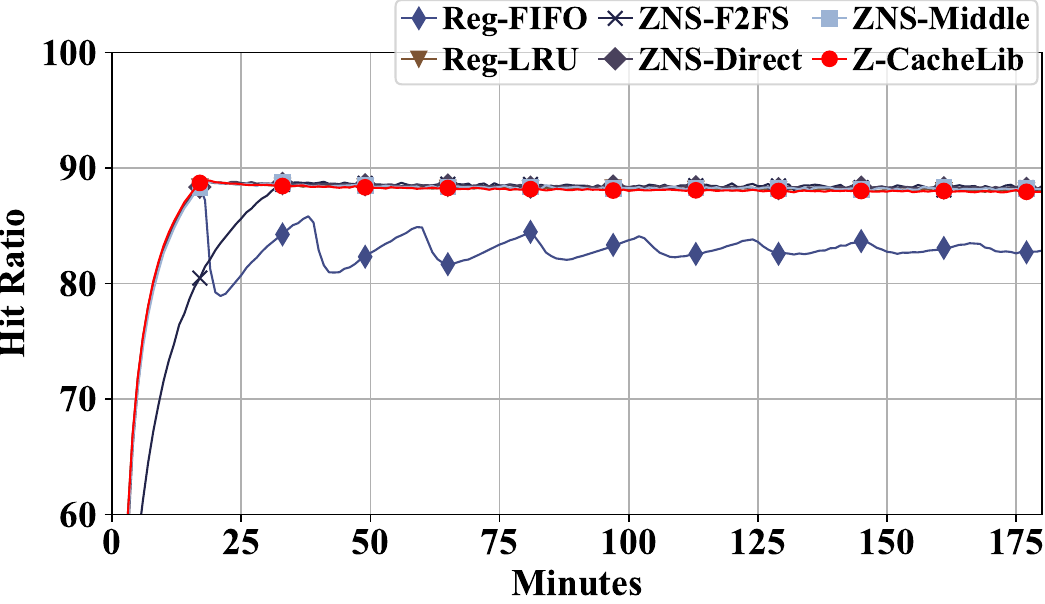}}
  \subfigure[\scriptsize{Cache hit ratio of \emph{\textbf{l2\_reg}}.}]{ 
  \label{fig:l2-reg-ratio} 
    \includegraphics[width=.32\linewidth]{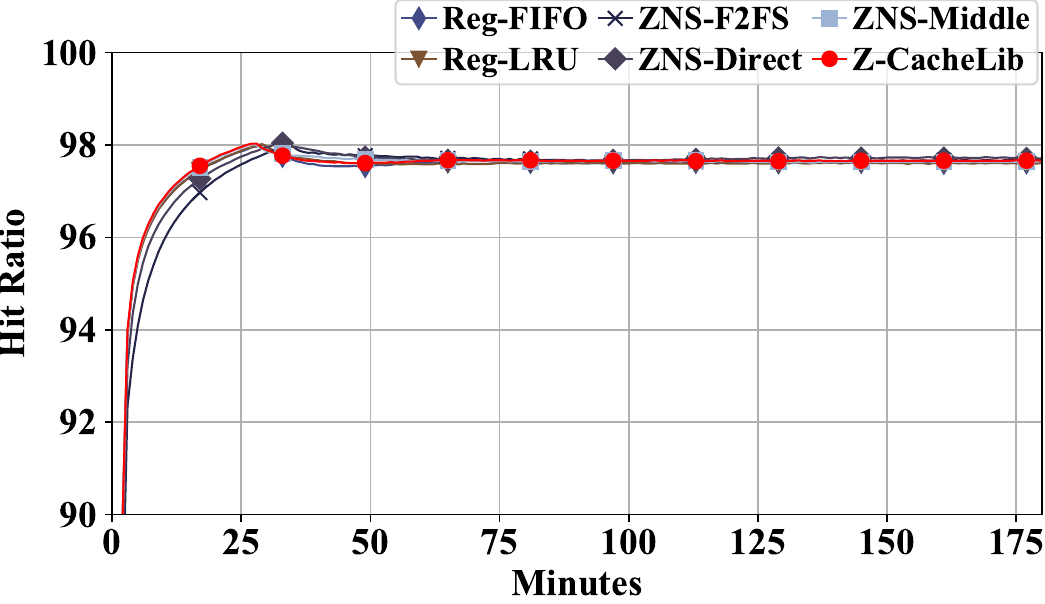}}
  \subfigure[\scriptsize{Cache hit ratio of \emph{\textbf{flat}}.}]{ 
  \label{fig:kvcache-ratio} 
    \includegraphics[width=.32\linewidth]{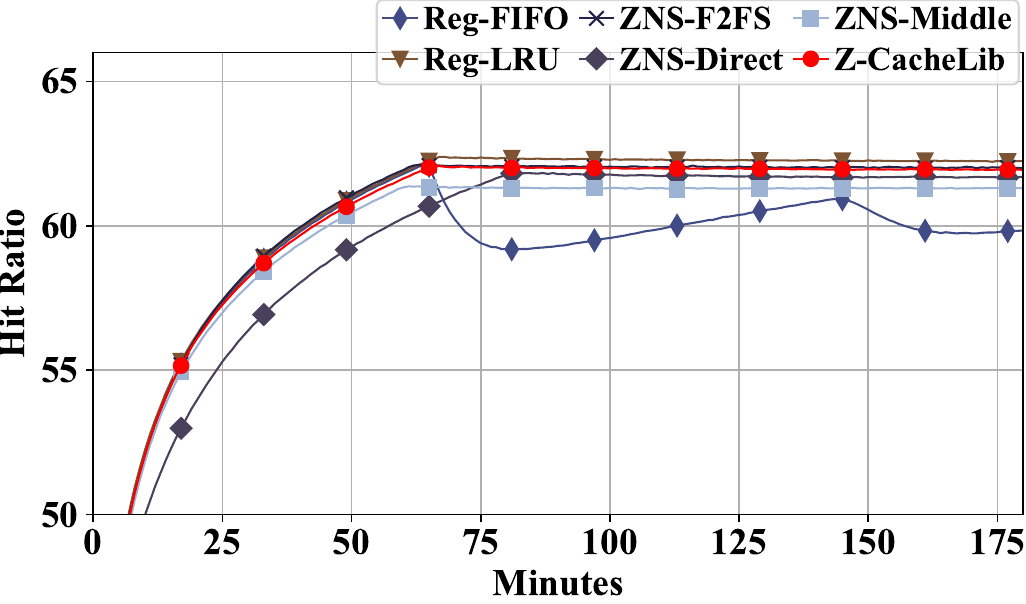}}

  \caption{Throughput (MOPM) variation and cache hit ratio comparison under different workloads.}
  \label{fig:overall} 
\end{figure*}

\subsection{Implementation Details}
We implemented ZNS-Middle and Z-CacheLib based on CacheLib V2022.09.05.00 and \emph{libzbd} v2.0.3. \emph{libzbd} is a user library that uses the kernel-provided zoned block device interface to achieve I/O and zone-based management including open, close, finish, and reset \cite{libzbd,bjorling2020zone}. The prototype will be open-sourced. It is tested on Ubuntu 22.04.

In Z-CacheLib, we use a tree-based map to implement the mapping from the region's virtual address directly to the ZNS physical address in zMap. In this way, we can perform a binary search to find the desired position to read. The reverse mapping of zMap is distributed and maintained by each zone and can be used to retrieve the status and information of valid regions. Storage Engine aligns with the same persistency level promised by CacheLib. zMap can be considered as part of CacheLib in-memory data structure. Since original CacheLib does not promise the data persistency before ``seal'' or ``close'' is called, we do not need to persist the zMap for each update. The zMap is persisted to the reserved first zone together with CacheLib metadata when CacheLib persists its own mapping and metadata information.

\subsection{Experimental Setup} \label{subsec:setup}
We conduct all the experiments on an ASUSTeK ESC4000 server with 2 Intel Xeon 4210 CPUs (24 Cores) and 192 GiB of memory. We use Western Digital Ultrastar DC ZN540 (1TB) ZNS SSD, which has 904 zones and 1077MiB zone capacity \cite{zn540}. To achieve a fair performance comparison with a compatible regular SSD, we use Western Digital Ultrastar DC SN540 (960GiB) which has the same hardware as ZN540. 

In order to evaluate CacheLib with F2FS on ZNS SSDs, we need to reserve a small random writable device for its metadata. To avoid the performance bottleneck from the random writable device, we use Linux null block device driver \cite{zbd-emulation} to emulate a memory-backing conventional block device that supports random writes. F2FS is a log-structured Linux file system designed to perform well on modern flash storage devices. The zoned block device support was added to F2FS with kernel 4.10 \cite{lee2015f2fs, zonef2fs}. The version we used in the evaluation is F2FS-tools 1.15.0. 

We implemented and deployed CacheLib on F2FS (ZNS-F2FS), CacheLib on the middle layer (ZNS-Middle), and directly zone support layer (ZNS-Direct) as the baselines on ZNS SSDs. At the same time, we also deployed CacheLib with different eviction policies (FIFO and LRU) on the raw regular SSDs (Reg-FIFO and Reg-LRU). They are compared with Z-CacheLib under the measurement matrics of CacheLib throughput \emph{million operations per minute} (MOPM), CacheLib \emph{hit ratio per minute}, and \emph{write amplification}. All the tests run at least 3 times and we present the average result. 
We evaluated the six schemes with three different workloads in CacheBench, which is a real-world cache benchmarking tool introduced in CacheLib \cite{berg2020cachelib, mcallister2021kangaroo, cachebench}. 

The three workloads are synthetically generated based on different cache traces at Meta \cite{cachebench}: \textbf{\emph{l2\_wc}} (60\% get ratio) and \textbf{\emph{l2\_reg}} (88\% get ratio) are the workloads from two different production cache pools with large objects (several KBs or even larger size); \textbf{\emph{flat}} (98.5\% get ratio) is from a cache for mixed small and large objects. All three workloads are mixed with cache lookup and cache insert. CacheBench provides the \emph{opDelayNs} configuration to measure the performance of workloads at a certain throughput and we set \emph{opDelayNs} to zero to generate the stressful workloads.

When running CacheBench to evaluate Cachelib, there are 3 stages. Stage 1, CacheLib is generating regions and filling up the reserved zones, and no eviction and GC will be triggered (filling-stage). The throughput usually varies a lot in this stage. State 3, all cache space is used, and CacheLib will reclaim cache regions for upcoming regions under the FIFO or LRU eviction policy. In this stage, GC will not be frequently triggered due to the OP space (evicting-stage). Stage 3, all zones are filled up and GC continues to reclaim the space for upcoming regions (stable-stage). We mainly focus on the throughput and cache hit ratio at the \textbf{stable-stage} in the following evaluations, since most of the time a CacheLib instance is running in the stable-stage.

\subsection{Overall Comparison} \label{sec:overall-evaluation}

In all testing cases, CacheLib uses 887 GiB size of virtual space for regions and we reserve 904 ZNS SSD zones (the OP ratio is about 7\%). We run the 3 workloads in CacheBench for more than 180 minutes to ensure they issue at least 50 million cache operations and 2000GiB cache writes. The run-time throughput variation and the run-time cache hit ratio (per minute) variation of three workloads are shown in Figure \ref{fig:overall}. We summarize the average stable-stage throughput and the average stable-stage cache hit ratio in Table \ref{T1}. In general, under the workloads of \emph{l2\_reg}, \emph{l2\_wc}, and \emph{flat}, Z-CacheLib can achieve 205\%, 519\%, and 142\% higher throughput than that of ZNS-F2FS respectively. The cache hit ratio is only reduced by about 0.24\%, 0.02\%, and 0.05\% respectively. Compared to Reg-LRU, we can also get 108\%, 27\%, and 8\% throughput improvement respectively. Compared to Reg-FIFO, we can also get 5.2\%, 0.04\%, and 2.42\% hit ratio improvement respectively. Overall, Z-CacheLib can achieve better throughput
 than Reg-FIFO while keeping the hit ratio similar to Reg-LRU.

\begin{table}[]
\centering
\caption{The stable-stage throughput and hit ratio.}
\label{T1}
\setlength{\tabcolsep}{3.5pt}
\begin{tabular}{@{}l|ccccll@{}}
\toprule
 \multirow{2}{*}{stable-stage}          & \multicolumn{3}{l}{Throughput (MOPM)} & \multicolumn{3}{l}{Hit Ratio (\%)} \\ 
           & \textbf{\textit{l2\_wc}} & \textbf{\textit{l2\_reg}} &  \multicolumn{1}{l}{\textbf{\textit{flat}}} & \textbf{\textit{l2\_wc }} & \textbf{\textit{l2\_reg}} & \textbf{\textit{flat}} \\ \midrule
Reg-FIFO   & 0.99   & 12.76   & \textbf{43.10}        & 83.02 & 97.62 & 59.58      \\
Reg-LRU    & 0.50   & 10.67   & 38.00        & 88.19 & 97.60 & \textbf{62.31}     \\
ZNS-F2FS   & 0.34   & 2.20    & 17.14        & 88.33 & 97.68 & 62.05     \\
ZNS-Direct & 0.60   & 4.83    & 22.36        & \textbf{88.39} & \textbf{97.70} & 61.77      \\
ZNS-Middle & 0.28   & 6.24    & 26.90        & 88.28 & 97.64 & 61.31      \\
Z-CacheLib & \textbf{1.04}   & \textbf{13.63}   & 41.58        & 88.09 & 97.66 & 62.00      \\ \bottomrule
\end{tabular}
\end{table}

In the stable-stage, only Z-CacheLib and Reg-FIFO can keep a high throughput, and other schemes drop quickly. For example, under the workload of \emph{l2\_wc}, the throughput of ZNS-Middle drops more than 70\%. Meanwhile, Reg-LRU and ZNS-Direct both dropped about 50\%. The ZNS-F2FS has the lowest throughput due to its heavy overhead from its filesystem design and it is lower than ZNS-Middle. It validates the effectiveness of removing the block interface, simplifying the mapping, using parallel writing, and applying our GC designs. The ZNS-Middle has lower throughput than Reg-LRU because of its large erase unit (i.e., 1077MiB zone). The ZNS-Direct can have higher throughput than Reg-LRU and it is a promising design when the zone size is small.

The cache hit ratios of LRU-based schemes are similar as shown in Figure \ref{fig:l2-reg-ratio}, \ref{fig:l2-wc-ratio}, and \ref{fig:kvcache-ratio}. Under the workload of \emph{l2\_wc}, The Reg-FIFO has an unstable and lower throughput due to its simple eviction policy.
In general, the evaluations validate that keeping the block interface and using a middle layer (ZNS-F2FS and ZNS-Middle) to adapt the flash cache on ZNS SSDs can introduce extra performance overhead and lead to explicit low throughput compared with Reg-SSD. Moreover, directly using the zone interface (ZNS-Direct) may incur lower throughput though it is a GC-free design. In the end, Z-CacheLib can achieve FIFO-level throughput and LRU-level hit ratio while keeping a very low WA (lower than 1.1). Compared to regular SSDs, Z-CacheLib successfully achieves higher throughput (up to 108\% improvement) compared with Reg-LRU. 

\subsection{Write Amplification} \label{sec:evaluation-wa}

In the previous overall comparison tests, we also collected the total amount of actual ZNS writes when the data written by CacheLib in each benchmarking experiment for ZNS-F2FS, ZNS-Middle, ZNS-Direct, and Z-CacheLib, to calculate the overall WA. The WA statistics are shown in Table \ref{T3} when CacheLib generates 1400GiB writes.

\begin{table}[h]
\vspace{-2px}
\centering
\caption{The write amplification factor summary.}
\label{T3}
\begin{tabular}{llll}
\toprule
          & \emph{\textbf{l2\_wc}} & \emph{\textbf{l2\_reg}} & \emph{\textbf{flat}} \\ \midrule
ZNS-F2FS     & 2.51     & 2.84   & 2.08    \\
ZNS-Middle  & 2.11     & 2.37   & 1.97    \\
ZNS-Direct & 1.00      & 1.00   & 1.00    \\
Z-CacheLib & 1.01     & 1.01   & 1.01    \\ \bottomrule
\end{tabular}
\vspace{-2px}
\end{table}

In ZNS-Middle and ZNS-F2FS, the size of the erase unit (zone) is much larger than regular SSDs, reducing the efficiency of GC and introducing larger WA. Therefore, they have similar WA factors. As provided by a filesystem, the ZNS-F2FS has much more overhead than the ZNS-Middle, so its WA factor is slightly higher than ZNS-Middle. Z-CacheLib and ZNS-Direct are GC-free design so their WA factor are very low.

In Z-CacheLib, since top-down eviction (zLRU) and bottom-up eviction (zDrop) reduce valid regions in victim zones, the WA is reduced to about 1, which is almost no WA happening. It verifies that Z-CacheLib successfully identifies the valid regions that have little influence on the cache hit ratio. Also, firstly evicting those regions or avoiding migrating those regions can benefit both performance and write amplification significantly.

\begin{figure}[!t]
\hspace{-8px}
  \centering 
  \subfigure[\scriptsize{Tradeoff of each component.}]{ 
  \label{fig:tradeoff-each-design} 
    \raisebox{2.2mm}{\includegraphics[width=.49\linewidth]{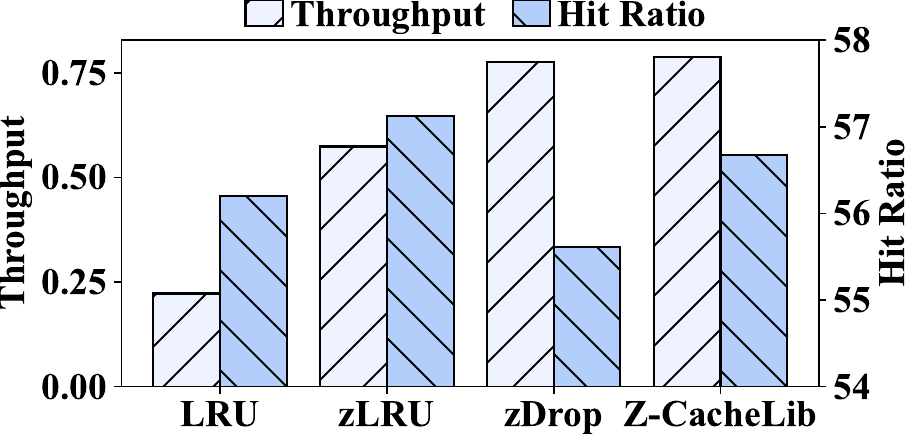}}} 
  \subfigure[\scriptsize{Tradeoff of vOP space.}]{ 
  \label{fig:tradeoff-vop} 
    \includegraphics[width=.50\linewidth]{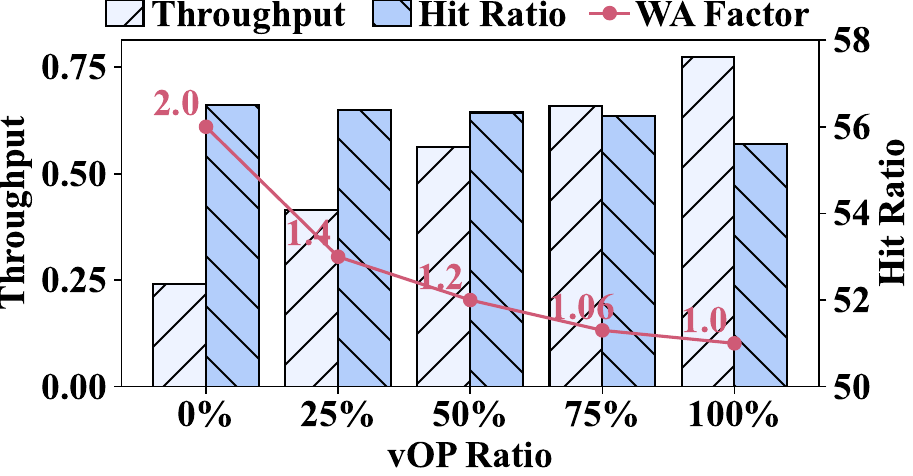}}
  \caption{The tradeoffs of throughput, hit ratio, and WA with zLRU, zDrop, and vOP.}
  \label{fig:tradeoff-all-zcache} 
\hspace{-8px}

 \end{figure}

\subsection{Tradeoff Analysis} \label{sec:tradeoff-analysis}

In this section, we first evaluated zLRU and zDrop separately to validate their individual contributions, as illustrated in Figure \ref{fig:tradeoff-each-design}. Moreover, we presented tradeoff evaluations under various OP ratios, cache sizes, and zone sizes to provide a better understanding of flash cache and Z-CacheLib.

\textbf{Tradeoff of each component} \label{sec:each-design}
We reserved 100 zones (1077 MiB) to build an 88 GiB cache and evaluated its performance using the \emph{l2\_wc} workload. The zLRU offers approximately 3X throughput compared to using LRU. The zLRU reorders its LRU list to effectively utilize the OP space. Therefore, the victim zones have smaller valid data size, leading to a smaller WA factor and higher throughput. Additionally, the hit ratio improves in this workload, rendering our eviction policy effective at finding valid regions to clean. While zDrop significantly improves throughput (by 38\% compared to using zLRU and 259\% compared to using LRU), its hit ratio is lower due to its aggressive eviction policy. Z-CacheLib combines zLRU and zDrop, resulting in both higher throughput and a higher hit ratio compared to LRU in our evaluation.

\textbf{Different vOP Ratios} The vOP ratio is the fraction of vOP space to the cache space (including vOP space). A higher vOP ratio will have a larger vOP space. We conducted experiments on 100 zones (13\% OP ratio) to analyze the influence of using different vOP ratios of zDrop, as shown in Figure \ref{fig:tradeoff-vop}. The vOP ratio is utilized by zDrop to clean regions. When the regions of victim zone are in vOP space, the zDrop will evict these regions to reduce data rewritten. Therefore, a higher vOP ratio improves throughput and the WA factor, but it results in a lower hit ratio. Thus, there is a trade-off between the hit ratio and the vOP ratio. However, in our overall evaluation, one zone size is considerably smaller than the cache size (only 1.1\%), so even if we set vOP to 100\%, the hit ratio does not decrease significantly.

\begin{figure}[!t]
\hspace{-8px}
  \centering 
  \subfigure[\scriptsize{Stable-stage throughput comparison.}]{ 
  \label{fig:overprovision-throughput} 
    \includegraphics[width=.50\linewidth]{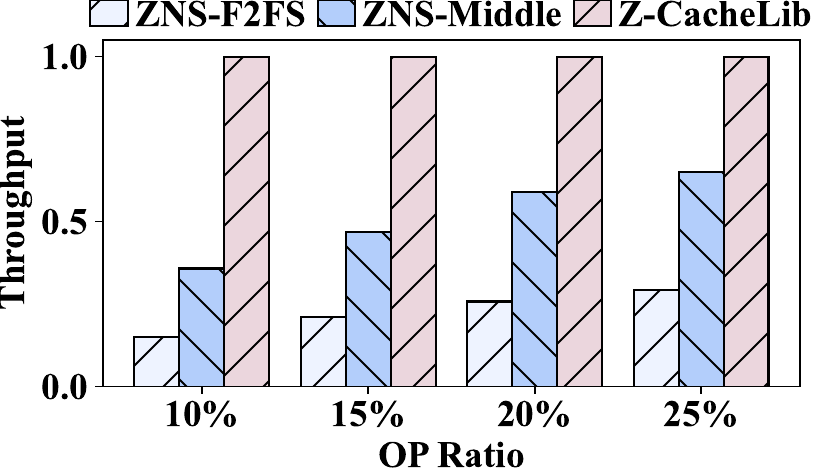}}
  \subfigure[\scriptsize{Write Amplification comparison.}]{ 
  \label{fig:overprovision-wa} 
    \includegraphics[width=.49\linewidth]{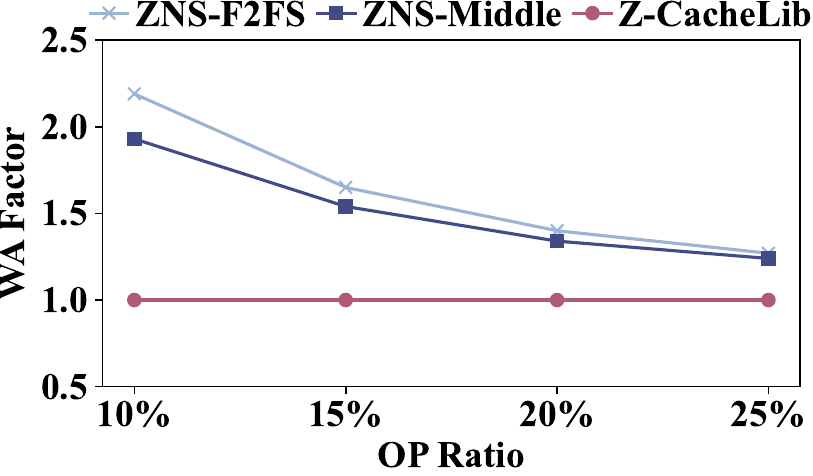}}
  \caption{The cache throughput and WA factor with different OP ratios.}
  \label{fig:overprovisionchange} 
  \hspace{-8px}
\vspace{-8px}
\end{figure}

\textbf{Different OP Ratios} As shown in Figure \ref{fig:overprovisionchange}, we conducted experiments with different OP ratios of 904 zones under \emph{l2\_wc} workload. When we increase the OP ratio, the cache size will be decreased accordingly. As shown in Figure \ref{fig:overprovision-throughput}, the throughput increases as the OP ratio increases in ZNS-F2FS, ZNS-Middle, and ZNS-Cache. For ZNS-F2FS, the throughput doubles as the OP ratio increases from 10\% to 25\%. In those three schemes, with more extra space, GC will have a lower performance impact on the foreground throughput. Differently, GC is not the performance bottleneck of Z-CacheLib due to cross-layer optimizations. The throughput of Z-CacheLib decreases slightly when we increase the OP ratio from 15\% to 25\%, which is caused by more cache misses with the smaller cache size. The cache hit ratios drop as OP increases (cache size decreases) in all four schemes. As expected in Figure \ref{fig:overprovision-wa}, the WA all decrease when we over-provision more space. In general, the WA of Z-CacheLib is not sensitive to the OP ratio. We can use a lower OP in Z-CacheLib for a higher cache hit ratio.

\begin{figure}[!t]
\hspace{-8px}
  \centering 
  \subfigure[\scriptsize{Stable-stage throughput comparison.}]{ 
  \label{fig:size-throughput} 
    \includegraphics[width=.49\linewidth]{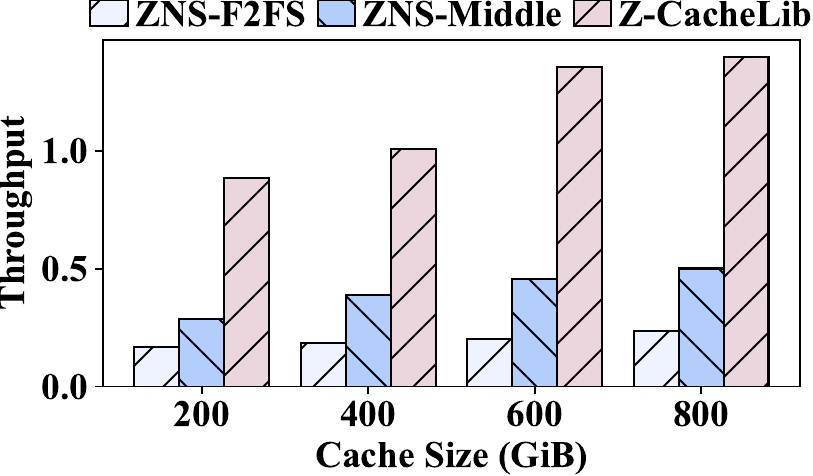}}
  \subfigure[\scriptsize{Hit ratio comparison.}]{ 
  \label{fig:size-ratio} 
    \includegraphics[width=.49\linewidth]{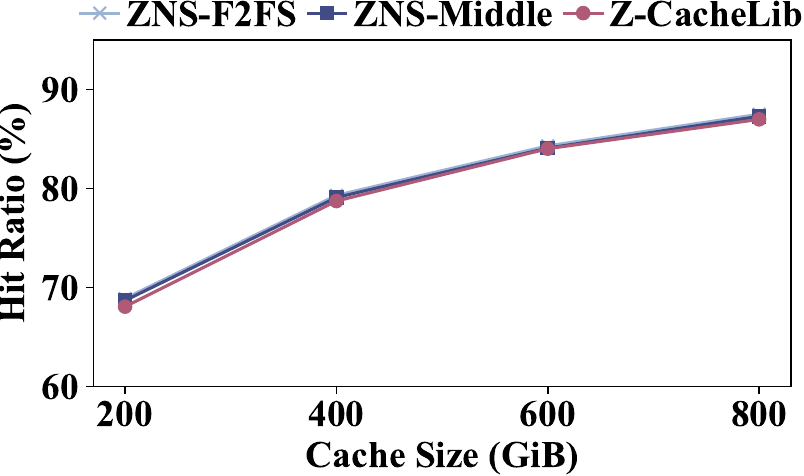}}
  \caption{The cache throughput and hit ratio with different cache sizes.}
  \label{fig:sizechange} 
  \hspace{-8px}
  \vspace{-12px}
\end{figure}

\textbf{Different Cache Sizes} We also evaluated the throughput, cache hit ratio, and WA variations with different ZNS spaces reserved for CacheLib as shown in Figure \ref{fig:sizechange}. We evaluate the 4 schemes with the ZNS different space sizes from 200 to 800 zones under the \emph{l2\_wc} workload. In each test, we write 2$X$ reserved space size of data to ZNS SSD. The OP ratio is 20\% in all schemes. As shown in Figure \ref{fig:size-throughput}, as the cache size increases, the throughput of all 4 schemes increases explicitly. Z-CacheLib can always achieve the best throughput due to cross-layer optimizations.

As shown in Figure \ref{fig:size-ratio}, the cache hit ratio increases from about 68\% to 86\% as the reserved ZNS space increases. In all cases, the cache hit ratio difference is smaller than 0.2\%. It validated that, the evicted valid regions selected by Z-CacheLib during GC have a very small influence on the overall cache hit ratio. The cache hit ratio is mainly decided by the reserved ZNS space size. Importantly, Z-CacheLib effectively reduces the WA and achieves almost 1.

\begin{figure}[!t]
\hspace{-8px}
  \centering 
  \subfigure[\scriptsize{Stable-stage throughput comparison.}]{ 
  \label{fig:zonesize-throughput} 
    \includegraphics[width=.49\linewidth]{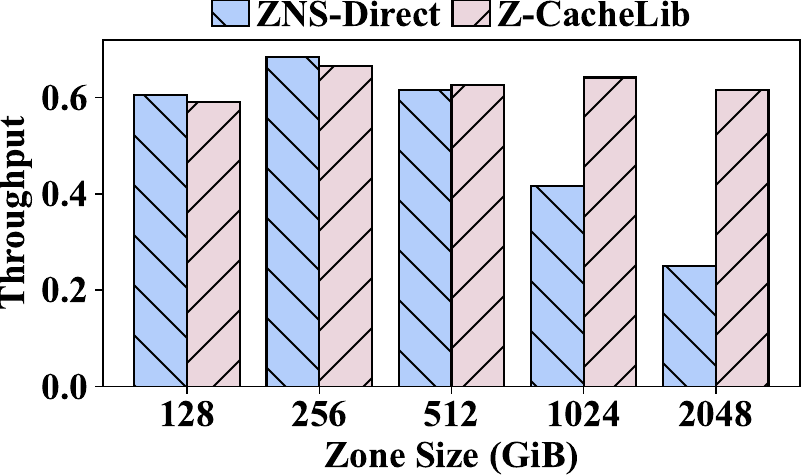}}
  \subfigure[\scriptsize{Hit ratio comparison.}]{ 
  \label{fig:zonesize-ratio} 
    \includegraphics[width=.49\linewidth]{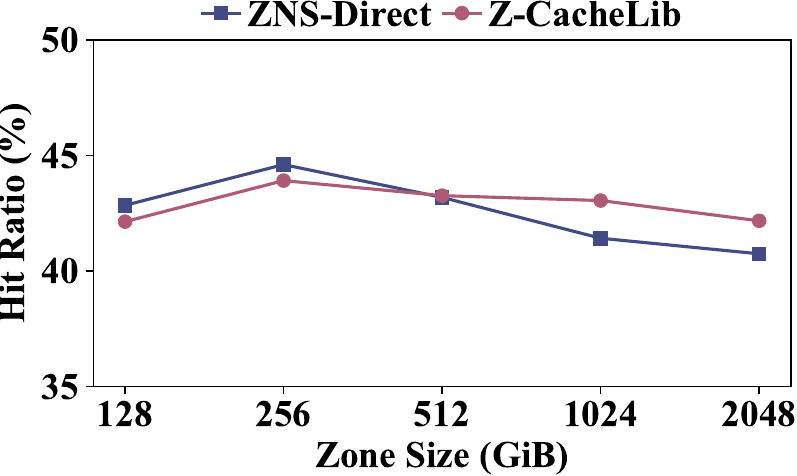}}
  \caption{The throughput under different zone sizes.}
  \label{fig:zonesize-change} 
  \hspace{-8px}
  \vspace{-4px}
\end{figure}

\textbf{Different Zone Sizes} We simulated 100 GiB ZNS SSDs with different zone sizes using a 1 TiB SN540 regular SSD with 900\% OP ratio to ensure stable write performance. Figure \ref{fig:zonesize-change} illustrates the throughput and hit ratios of Z-CacheLib and Z-Direct under different zone sizes. The simulated zones have the same I/O bandwidth regardless of their sizes. The throughput of Z-CacheLib can remain stable and exceed that of ZNS-Direct by up to 120\%. ZNS-Direct's performance will degrade further as the zone size exceeds 256 MiB as the inefficient management for large region size discussed in Section \ref{sec:exsiting}. When the zone size is small (e.g., 128 MiB), ZNS-Direct appears to be a favorable scheme. However, smaller zones may exhibit lower throughput. For instance, in eZNS\cite{min2023ezns}, a 96 MB zone size only achieves a 40 MiB/s write bandwidth, whereas a 1077 MiB zone can deliver a guaranteed 1000 MiB/s write bandwidth. Larger zones, which provide higher single-thread throughput, require minimal parallel control considerations but may entail more management overhead to handle their large size. Consequently, different zone sizes offer varying benefits. Fortunately, Z-CacheLib can manage ZNS SSDs with varying zone sizes.

\section{Conclusion}
In this paper, we introduce Z-CacheLib, a zoned storage optimized flash cache designed for ZNS SSDs. Z-CacheLib utilizes the zStorage Engine to facilitate I/O operations directly between cache data and ZNS SSDs, eliminating the overhead of block interfaces and complex mappings. Additionally, we introduce virtual over-provisioning space, zLRU, and zDrop mechanisms to achieve a balanced trade-off between cache throughput, cache hit ratio, and write amplification. Compared to regular SSDs, Z-CacheLib can achieve FIFO-level performance while maintaining an LRU-level hit ratio, showing promising benefits of using ZNS SSDs. 
Exploring workload-adaptive GC and cache administration on Z-CacheLib is our future work.
\newpage


\end{document}